\documentclass[
 reprint,
 amsmath,amssymb,
 aps,
 prl,
 superscriptaddress
]{revtex4-1}

\usepackage{graphicx}
\usepackage{dcolumn}
\usepackage{bm}
\usepackage{color}

\begin{document}

\title{Reentrant $\nu=1$ quantum Hall state in a two-dimensional hole system}

\author{A.L.\ Graninger}
\author{D.\ Kamburov}
\author{M.\ Shayegan}
\author{L.N.\ Pfeiffer}
\author{K.W.\ West}
\author{K.W.\ Baldwin}
\affiliation{Department of Electrical Engineering, Princeton University, Princeton, New Jersey 08544, USA}

\author{R.\ Winkler}
\affiliation{Department of Physics, Northern Illinois University, DeKalb, Illinois 60115, USA}
\affiliation{Materials Science Division, Argonne National Laboratory, Argonne, Illinois 60439, USA}

\date{\today}

\begin{abstract}
We report the observation of a reentrant quantum Hall state at the Landau level filling factor $\nu=1$ in a two-dimensional hole system confined to a 35-nm-wide (001) GaAs quantum well.  The reentrant behavior is characterized by a weakening and eventual collapse of the $\nu=1$ quantum Hall state in the presence of a parallel magnetic field component $B_{||}$, followed by a strengthening and reemergence as $B_{||}$ is further increased.  The robustness of the $\nu=1$ quantum Hall state during the transition depends strongly on the charge distribution symmetry of the quantum well, while the magnitude of $B_{||}$ needed to invoke the transition increases with the total density of the system.

\end{abstract}


\maketitle

Studies of two-dimensional electron systems (2DESs) confined to wide quantum wells (WQWs) have revealed a rich variety of quantum phases which are not seen in standard, single-layer heterojunctions or narrow QWs, including unique fractional quantum Hall states (QHSs)  \cite{Suen.PRL.92,Manoharan.PRL.97,Luhman.PRL.08,Shabani.PRL.10} and a bilayer Wigner solid  \cite{Manoharan.PRL.96}.  In these systems, the large QW width coupled with the electrostatic repulsion of the electrons results in a bilayer-like charge distribution which introduces an additional (layer) degree of freedom.  This extra degree of freedom impacts the 2DES electronic level structure as well as many-body interactions arising from the additional interlayer correlations, both of which can be tuned in a single WQW by changing the total density and/or the symmetry of the charge distribution.  Here we report on magnetotransport measurements in a two-dimensional {\it hole} system (2DHS) confined to a WQW.  2DHSs differ from their 2D electron counterparts in several aspects. Most notably, the 2D holes' $j=3/2$ effective spin (rather than $j=1/2$) and their strong spin-orbit coupling result in a non-parabolic energy band structure at zero magnetic field, and Landau levels that typically exhibit a non-linear dependence on perpendicular magnetic field \cite{Winkler.03}.  These unique subtleties of single-layer 2DHSs motivated the work we present here, where the holes are confined to a WQW and typically occupy two electric subbands.

Our measurements reveal a striking observation, namely, a reentrant integer QHS at the Landau level filling factor $\nu=1$ in the presence of a parallel magnetic field component $B_{||}$.  This phenomenon is characterized by a weakening and eventual collapse of the $\nu=1$ QHS as the 2DHS is tilted in field. The state is absent over a range of $B_{||}$, but at a higher $B_{||}$ it reemerges and strengthens with increasing $B_{||}$.  This reentrant behavior has no counterpart in 2DESs confined to WQWs, where the collapse of the $\nu=1$ QHS with increasing $B_{||}$ is permanent \cite{Lay.PRB.97,Manoharan.PRL.97}.  Our magnetotransport data also reveal clear dependencies on both the charge distribution symmetry and total density of the 2DHS. As the charge distribution is made increasingly asymmetric, the $\nu=1$ QHS becomes more robust during the transition, and as the density is increased, the range of $B_{||}$ over which the transition occurs shifts to higher $B_{||}$.

We studied a 2DHS confined to a 35-nm-wide GaAs QW grown via molecular beam epitaxy on a (001) GaAs substrate.  The WQW is surrounded on both sides by 95-nm-thick Al$_{0.24}$Ga$_{0.76}$As spacer layers and C $\delta$-doped layers.  Two samples were studied here:  Sample A, which was fitted with a Ti/Au front gate and an In back gate, and sample B, with just a front gate.  The gates allow us to change the density of the 2DHS, and in sample A we use the two gates to additionally tune the charge distribution symmetry, which we define as the difference in density between the front and back of the WQW as a fraction of the total density, $\delta p/p$.  As a result of the symmetric doping profile, the 2DHS is nearly balanced ($\delta p/p=0$) at zero gate bias.  To determine the balanced point more precisely, we carried out measurements at a fixed density for various $\delta p/p$ and judged the balanced point by identifying symmetric features in the magnetotransport data taken at both $\delta p/p > 0$ and $\delta p/p < 0$.  The zero-bias density and low-temperature mobility of both samples are $p = 1.7 \times 10^{11}$ cm$^{-2}$ and $\mu = 3 \times 10^{5}$  cm$^2$/Vs, respectively.  The samples were measured in a $^3$He refrigerator with a base temperature of $T\simeq$ 300 mK in which they were mounted on a rotating stage that allows us to change \emph{in situ} the tilt angle $\theta$ between an external magnetic field and the direction normal to the plane of the 2DHS.  The samples were fabricated in the van der Pauw geometry and magnetotransport data were obtained using standard low-frequency lock-in techniques.

\begin{figure}[ht]
\includegraphics[scale=0.34]{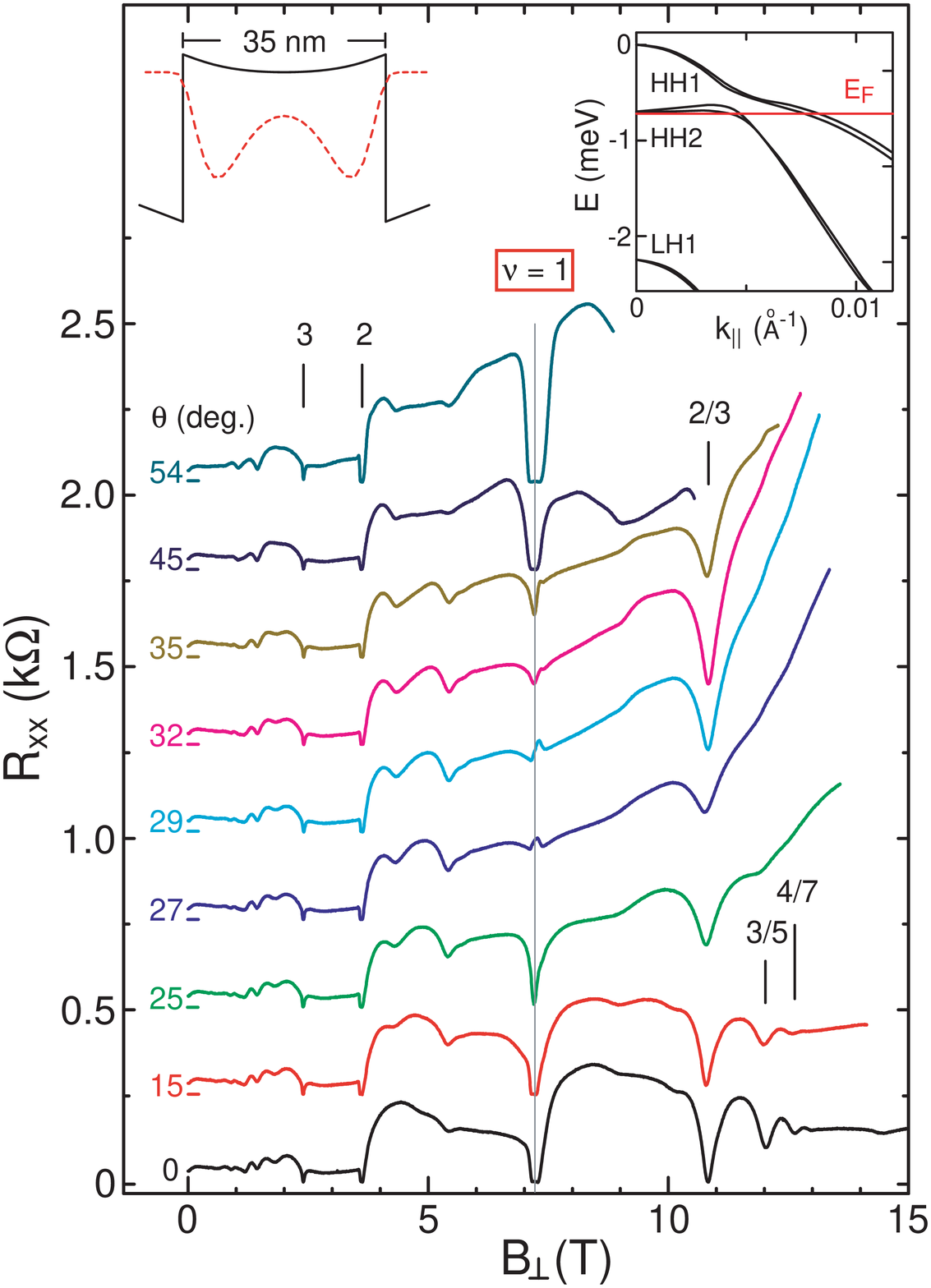}
\caption{\label{fig:fig1} (color online).  Magnetotransport data for sample A at a fixed density of $p = 1.74 \times 10^{11}$ cm$^{-2}$ in the balanced ($\delta p/p=0$) case as a function of $B_\perp$ for various $\theta$.  At low $\theta$, the $\nu=1$ QHS is well-developed.  As $\theta$ is increased the $\nu=1$ QHS weakens and eventually collapses at $\theta\simeq27^\circ$, and then reemerges and strengthens as $\theta$ is further increased.  During the $\nu=1$ transition, the fractional QHSs in the extreme quantum limit ($\nu=2/3$, $3/5$, and $4/7$) are also weakened.  Each trace is shifted vertically in $R_{xx}$ by $\sim$ 0.25 k$\Omega$; $R_{xx}=0$ for each trace is marked at $B_\perp=0$.  Left and right insets:  Self-consistently calculated potential profile and charge distribution, and $E$ vs.\ $k_{||}$ dispersion along $[100]$ showing two-subband occupation.}
\end{figure}

\begin{figure}[ht]
\includegraphics[scale=0.34]{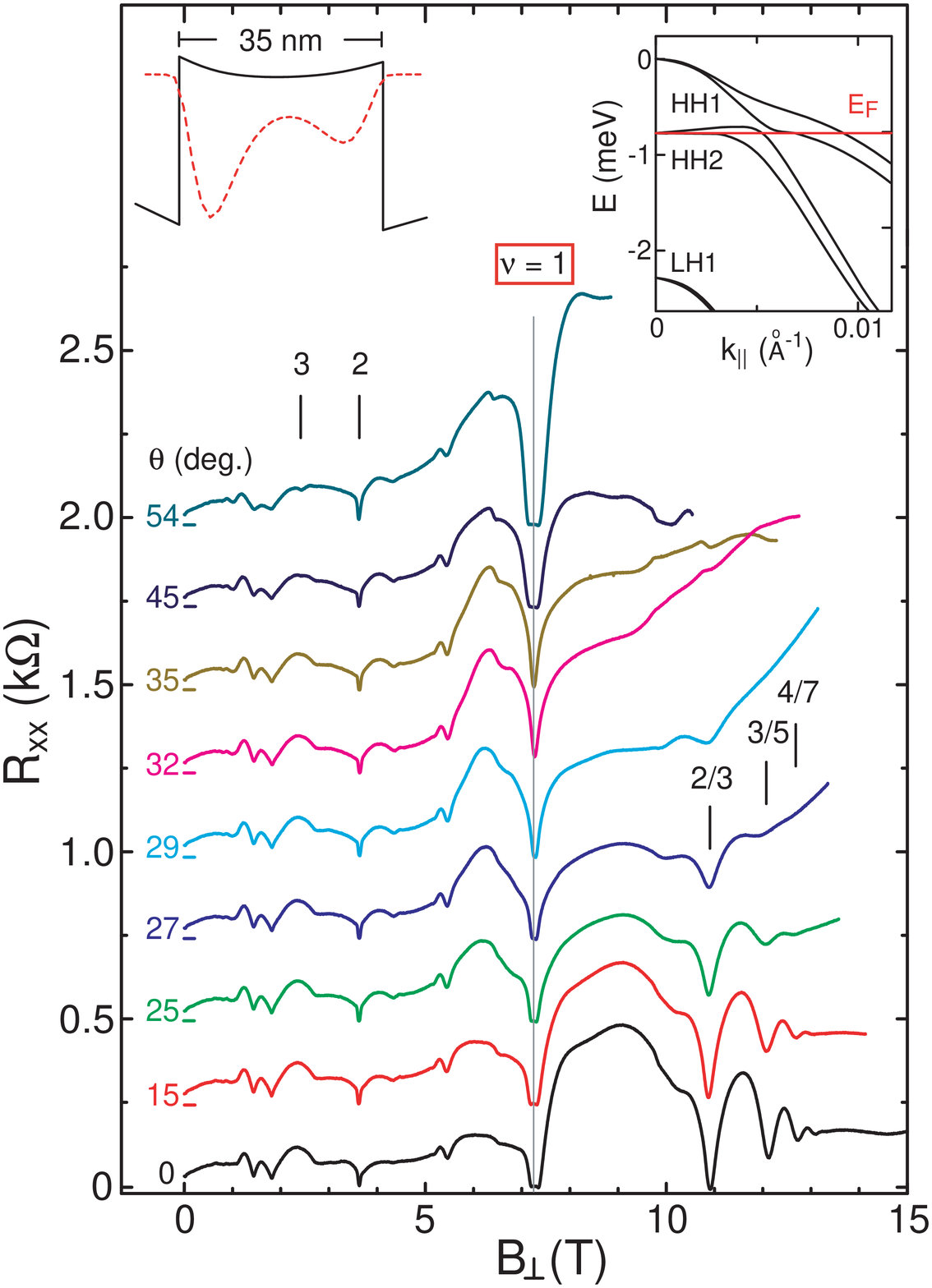}
\caption{\label{fig:fig2} (color online).  Magnetotransport data for sample A at $p = 1.74 \times 10^{11}$ cm$^{-2}$ in the imbalanced case where $\delta p/p=25\%$.  The data are qualitatively similar to the data of Fig.~\ref{fig:fig1} except that the $\nu=1$ QHS weakens slightly but does not collapse as in the $\delta p/p=0$ case.  Despite the robustness of the $\nu=1$ QHS here, the fractional QHSs in the extreme quantum limit are still severely affected over a similar range of $\theta$.  Left and right insets:  Self-consistently calculated potential profile and charge distribution, and $E$ vs.\ $k_{||}$ dispersion along $[100]$.}
\end{figure}

\begin{figure}[ht]
\includegraphics[scale=0.4]{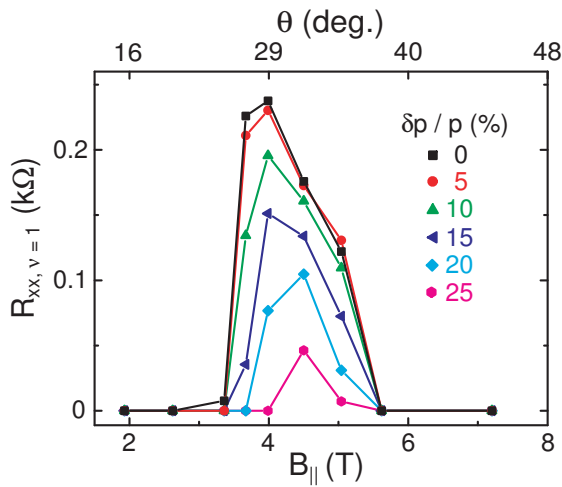}
\caption{\label{fig:fig3} (color online).  The value of $R_{xx}$ at $\nu=1$ vs.\ $B_{||}$ at $p = 1.74 \times 10^{11}$ cm$^{-2}$ for various $\delta p/p$.  As $|\delta p/p|$ is increased, the $\nu=1$ QHS becomes more robust over the range of $B_{||}$ in which the transition occurs.}
\end{figure}

\begin{figure}[ht]
\includegraphics[scale=0.32]{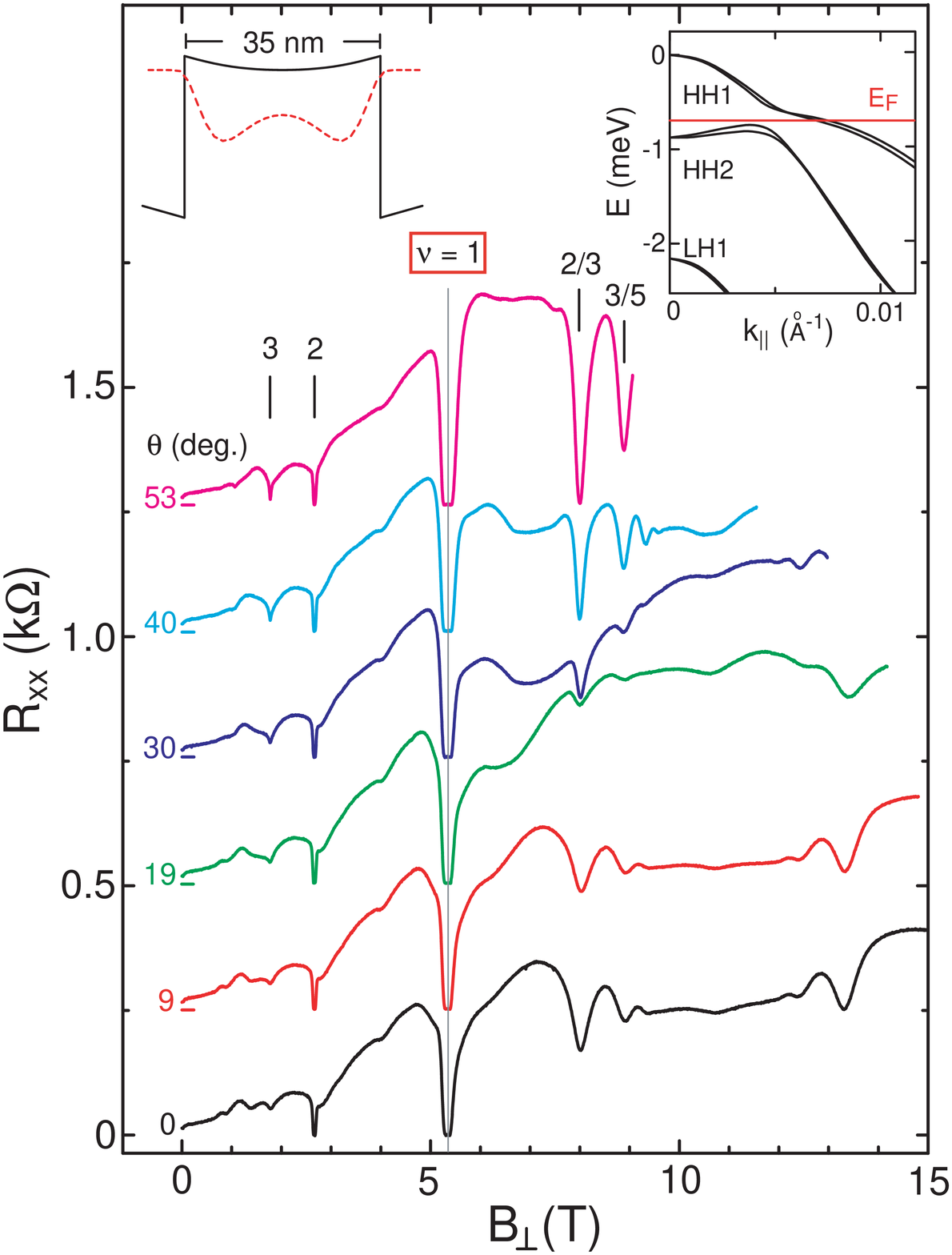}
\caption{\label{fig:fig4} (color online).  Magnetotransport data for sample A at a fixed density of $p = 1.29 \times 10^{11}$ cm$^{-2}$ for $\delta p/p=0$ for various $\theta$.  At $\theta=0$, there is a strong $\nu=1$ QHS which, in contrast to Fig.~\ref{fig:fig1}, monotonically strengthens with increasing $\theta$.  The fractional QHSs in the extreme quantum limit ($\nu=2/3$, $3/5$, and $4/7$) are weakened with increasing $\theta$; further increasing $\theta$ eventually strengthens these states to a point where at $\theta=53^\circ$ the $R_{xx}$ minima are deeper than at $\theta=0$.  Left and right insets:  Self-consistently calculated potential profile and charge distribution, and $E$ vs.\ $k_{||}$ dispersion along $[100]$ showing one-subband occupation.}
\end{figure}

\begin{figure}[ht]
\includegraphics[scale=0.4]{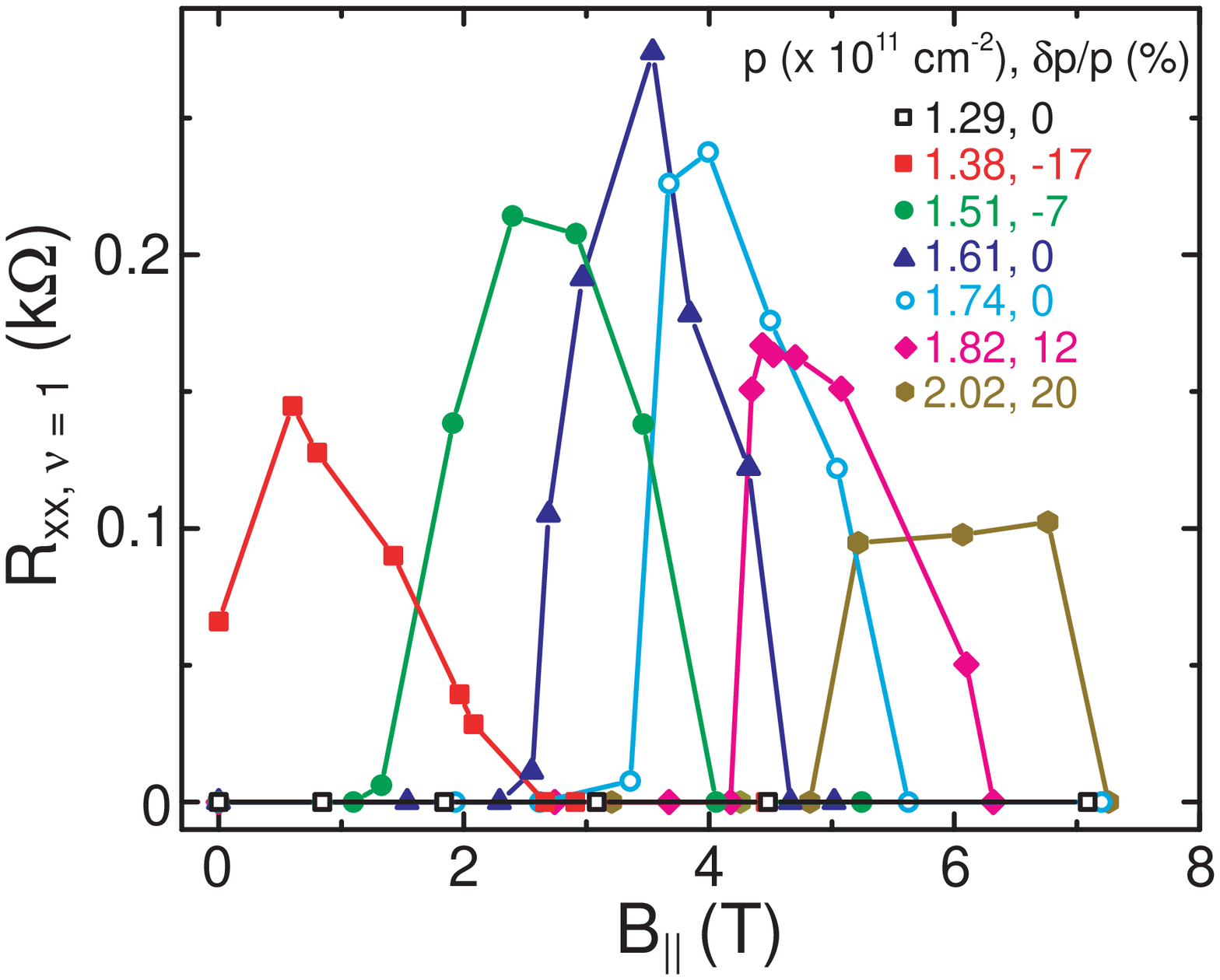}
\caption{\label{fig:fig5} (color online).  The value of $R_{xx}$ at $\nu=1$ vs.\ $B_{||}$ for various $p$; in each case the value of $\delta p/p$ is indicated.  As $p$ is decreased, the range of $B_{||}$ in which the $\nu=1$ transition occurs shifts to lower $B_{||}$.  At $p = 1.38 \times 10^{11}$ cm$^{-2}$, $R_{xx}$ at $\nu=1$ is already lifted up at $B_{||}=0$, while at $p = 1.29 \times 10^{11}$ cm$^{-2}$, no transition occurs.  Data obtained from sample A (B) are indicated by open (filled) symbols.}
\end{figure}

Figure~\ref{fig:fig1} shows the longitudinal resistance $R_{xx}$ as a function of the perpendicular component of the magnetic field $B_\perp$ for sample A at $p = 1.74 \times 10^{11}$ cm$^{-2}$ for various $\theta$.  The charge distribution is symmetric for all traces.  At $\theta=0$, there is a strong QHS at $\nu=1$ as evidenced by a wide and deep minimum in $R_{xx}$.  As $\theta$ is increased, the minimum becomes narrower, and by $\theta=27^\circ$ it completely collapses.  In this phase, what remains at $\nu=1$ is simply a doublet feature with two shallow minima at the background value of $R_{xx}$.  Further increasing $\theta$ results in a remarkable reemergence of the $\nu=1$ QHS; by $\theta=35^\circ$ a minimum in $R_{xx}$ is developing, and at $\theta=45^\circ$ a strong QHS at $\nu=1$ has set in.  Increasing $\theta$ beyond this point strengthens the $\nu=1$ QHS with no apparent indication of another transition.  Figure~\ref{fig:fig1} also displays the results of self-consistent calculations (for $B=0$) using the full $8\times8$ Kane Hamiltonian \cite{Winkler.03}. The left inset shows the potential profile and charge distribution, exhibiting the expected bilayer-like nature of our 2DHS.  The right inset shows the energy band dispersion along the $[100]$ direction indicating that two electric subbands of the heavy-hole band (HH1 and HH2) are occupied at this density.  Note that the energies of each of these subbands are split at finite wave vectors $k_{||}$ because of the spin-orbit interaction.

The role of charge distribution symmetry is illustrated in Fig.~\ref{fig:fig2}, which shows $R_{xx}$ vs.\ $B_\perp$ for the same density and the same values of $\theta$ as in Fig.~\ref{fig:fig1}.  However, in Fig.~\ref{fig:fig2} the asymmetry of the charge distribution is  $\delta p/p=25\%$.  In this case, at $\theta=0$ there is still a strong QHS at $\nu=1$.  With increasing $\theta$ , the $\nu=1$ QHS weakens over a similar range of $\theta$ as in the balanced case of $\delta p/p=0$ (Fig.~\ref{fig:fig1}), but it never collapses, as evidenced by the sustained deep $R_{xx}$ minimum at $\nu=1$ during the transition.  Similar to the balanced case, as $\theta$ is further increased beyond this range, the $\nu=1$ QHS strengthens.

The robustness of the $\nu=1$ QHS during this tilt-induced transition as a function of charge distribution asymmetry is illustrated in Fig.~\ref{fig:fig3}.  Here, the value of $R_{xx}$ at $\nu=1$ at $p = 1.74 \times 10^{11}$ cm$^{-2}$ is plotted vs.\ $B_{||}$ and $\theta$ for a number of $\delta p/p$.  A higher value of $R_{xx}$ here corresponds to a weaker $\nu=1$ QHS.  Figure~\ref{fig:fig3} demonstrates the primary effect of increased charge distribution asymmetry:  As $\delta p/p$ is increased, the $\nu=1$ QHS monotonically strengthens in the midst of the transition from a collapsed state at $\delta p/p=0$ to a QHS with a deep $R_{xx}$ minimum at $\delta p/p=25\%$.  The data also indicate that as $\delta p/p$ is increased, the critical values of $B_{||}$ and $\theta$ near which the transition occurs increase slightly.  We note that data for negative $\delta p/p$, omitted for clarity, show the same monotonic strengthening of the $\nu=1$ QHS.

To further characterize the nature of this transition, we obtained magnetotransport data for various densities.  Figure~\ref{fig:fig4} displays $R_{xx}$ vs.\ $B_\perp$ for $p = 1.29 \times 10^{11}$ cm$^{-2}$ and $\delta p/p=0$ for different values of $\theta$.  Here the $\nu=1$ QHS, which is well-developed at $\theta=0$, does not go through a transition as in the case of $p = 1.74 \times 10^{11}$ cm$^{-2}$.  Instead, it monotonically strengthens with increasing $\theta$ and through $\theta=53^\circ$ shows no signs of weakening.  As seen in the insets of Fig.~\ref{fig:fig4}, the calculated charge distribution in this case is distinctly less bilayer-like, resembling more of a thick 2DHS, while the $E$ vs.\ $k_{||}$ dispersion predicts the occupation of just the HH1 subband.

Figure~\ref{fig:fig5} summarizes the behavior of the $\nu=1$ transition at different densities.  The data in Fig.~\ref{fig:fig5} were obtained from both samples A and B, and for sample B, where the change in density was induced solely by a front gate, the associated $\delta p/p$ for each density is indicated.  The data show a clear dependence of the critical range of $B_{||}$ where the $\nu=1$ QHS weakens on the density of the 2DHS.  At the highest density of $p = 2.02 \times 10^{11}$ cm$^{-2}$, the critical range is centered about $B_{||}\simeq 6$ T.  As $p$ is decreased, the range monotonically shifts to lower $B_{||}$.  At $p = 1.38 \times 10^{11}$ cm$^{-2}$, $R_{xx}$ at $\nu=1$ is already lifted up at $B_{||}=0$, and increasing $B_{||}$ eventually strengthens the $\nu=1$ QHS as in the higher density cases.  Further decreasing the density to $p = 1.29 \times 10^{11}$ cm$^{-2}$ abruptly concludes this trend with a strong $\nu=1$ QHS at $B_{||}=0$ that only strengthens with tilt (see Fig.~\ref{fig:fig4}).

Before turning to a discussion of the $\nu=1$ transition, we briefly mention one other interesting feature exhibited by the magnetotransport data for $p = 1.74 \times 10^{11}$ cm$^{-2}$.  For both $\delta p/p=0$ (Fig.~\ref{fig:fig1}) and $\delta p/p=25\%$ (Fig.~\ref{fig:fig2}), the data exhibit a well-developed sequence of fractional QHSs at the filling factors $\nu=2/3$, $3/5$, and $4/7$.  When $\theta$ is increased such that the $\nu=1$ QHS is weakened, these fractional QHSs also go through a similar weakening.  In both cases, the QHSs at $\nu=3/5$ and $4/7$ are destroyed by a monotonic increasing of $R_{xx}$ with $B_\perp$ beyond $\nu=1$.  The $\nu=2/3$ QHS is also affected, but appears to remain stronger in the balanced case.  Interestingly, the data for $p = 1.29 \times 10^{11}$ cm$^{-2}$ (Fig.~\ref{fig:fig4}) also exhibit a weakening of these fractional QHSs at similar values of $\theta$, despite the absence of the $\nu=1$ transition at this density.  As $\theta$ is further increased, however, these fractional QHSs reemerge and by $\theta=53^\circ$ appear to be stronger than at $\theta=0$.  Because of the limited total magnetic field, we were not able to fully investigate these fractional QHSs at high $\theta$ in the higher density cases.

We now turn to possible origins of the observed $\nu=1$ transition.  In a standard, single-layer 2DES confined to a narrow GaAs QW or a heterojunction, the $\nu=1$ QHS arises from the Fermi level $E_F$ residing in the Zeeman energy gap which separates the lowest set of spin-split Landau levels (LLs). This QHS becomes stronger with the application of $B_{||}$ since $B_{||}$ increases the Zeeman energy.  On the other hand, in a bilayer-like 2DES, when the energy separation $\Delta$ between the first two electric subbands is smaller than both the Zeeman and cyclotron energies, the $\nu=1$ QHS corresponds to $E_F$ lying in the $\Delta$ energy gap. It is well known that $\Delta$ is reduced by an applied $B_{||}$ in bilayer-like 2DESs \cite{Hu.PRB.92}, and experiments in 2DESs confined to either WQWs \cite{Lay.PRB.97, Manoharan.PRL.97} or double QWs \cite{Murphy.PRL.94} have indeed shown that the application of a sufficiently strong $B_{||}$ leads to the collapse of the $\nu=1$ QHS into a compressible state. However, the collapse of $\Delta$ with increasing $B_{||}$ as seen in bilayer-like 2DESs cannot explain the \emph{reemergence} of the $\nu=1$ QHS at higher $B_{||}$ that we observe in our 2DHS.

While the reentrant $\nu=1$ QHS we observe has no counterpart in 2DESs, there has been a report of an anomalous, small magnetoresistance peak near the $\nu=1$ $R_{xx}$ minimum in a 2DHS confined to a (110) GaAs/Al$_x$Ga$_{1-x}$As heterojunction \cite{Fischer.PRB.07}.  In their experiments, which are done only in perpendicular magnetic fields, Fischer \emph{et al.}\ observe a slight weakening, but not a complete collapse, of the $\nu=1$ QHS.  Corroborating their data with the results of LL dispersion calculations, they attribute their observation to an anticrossing of spin-up and spin-down LLs.  It is possible that the origin of the reentrant $\nu=1$ QHS we observe is also a crossing of LLs with different spin and/or subband indices.  Unfortunately, the confirmation of such a scenario by theoretical calculations is inhibited by the difficulty of calculating the LL dispersions for a bilayer-like 2DHS in the presence of a parallel magnetic field component {\footnote{Both the multi-band structure of 2D holes and a tilted field geometry imply complicated couplings between different Landau harmonic oscillators.  A joint treatment of both effects would be exorbitantly complex to implement and also demanding when solved numerically.}.  However, it is very likely that the second electric subband plays a major role in the $\nu=1$ transition, given the evolution and eventual disappearance of this phenomenon as the 2DHS is tuned from a bilayer-like system with expected two-subband occupation to an increasingly single-layer-like system with one-subband occupation.  This conjecture is also consistent with previous reports on single-subband 2DHSs in tilted magnetic fields which do not show a $\nu=1$ transition with increasing $\theta$ \cite{Davies.PRB.91,Muraki.PRB.99,Pan.PRB.05}.  The general complexity of the LL structure of 2DHSs can be seen in Fig. 3 of Ref.\ \cite{Muraki.PRB.99} which shows the LL dispersions (at $\theta=0$) for a narrow GaAs QW. The dispersions are highly non-linear and despite the much larger value of $\Delta$, LLs of the HH2 subband are already interacting with those of HH1 \cite{Muraki.PRB.99}. This interaction is surely more dramatic in our system where $\Delta$ is smaller because of the larger QW width, and could lead to a crossing of HH1 and HH2 LLs and a reentrant $\nu=1$ QHS. 

The reentrant $\nu=1$ QHS we report here adds another twist to the subtle physics of 2DHSs in tilted magnetic fields.  While the origin of this phenomenon remains unknown, our data provide a number of clues, including dependencies on charge distribution symmetry and density, which we hope will stimulate further experimental and theoretical work.

\begin{acknowledgments}
We acknowledge support through the DOE BES (DE-FG02-00-ER45841) for measurements, and the Moore Foundation and the NSF (ECCS-1001719, DMR-0904117, and MRSEC DMR-0819860) for sample fabrication and characterization. Work at Argonne National Laboratory was supported by the DOE BES (DE-AC02-06-CH11357).
\end{acknowledgments}


\begin{thebibliography}{14}%
\makeatletter
\providecommand \@ifxundefined [1]{%
 \@ifx{#1\undefined}
}%
\providecommand \@ifnum [1]{%
 \ifnum #1\expandafter \@firstoftwo
 \else \expandafter \@secondoftwo
 \fi
}%
\providecommand \@ifx [1]{%
 \ifx #1\expandafter \@firstoftwo
 \else \expandafter \@secondoftwo
 \fi
}%
\providecommand \natexlab [1]{#1}%
\providecommand \enquote  [1]{``#1''}%
\providecommand \bibnamefont  [1]{#1}%
\providecommand \bibfnamefont [1]{#1}%
\providecommand \citenamefont [1]{#1}%
\providecommand \href@noop [0]{\@secondoftwo}%
\providecommand \href [0]{\begingroup \@sanitize@url \@href}%
\providecommand \@href[1]{\@@startlink{#1}\@@href}%
\providecommand \@@href[1]{\endgroup#1\@@endlink}%
\providecommand \@sanitize@url [0]{\catcode `\\12\catcode `\$12\catcode
  `\&12\catcode `\#12\catcode `\^12\catcode `\_12\catcode `\%12\relax}%
\providecommand \@@startlink[1]{}%
\providecommand \@@endlink[0]{}%
\providecommand \url  [0]{\begingroup\@sanitize@url \@url }%
\providecommand \@url [1]{\endgroup\@href {#1}{\urlprefix }}%
\providecommand \urlprefix  [0]{URL }%
\providecommand \Eprint [0]{\href }%
\providecommand \doibase [0]{http://dx.doi.org/}%
\providecommand \selectlanguage [0]{\@gobble}%
\providecommand \bibinfo  [0]{\@secondoftwo}%
\providecommand \bibfield  [0]{\@secondoftwo}%
\providecommand \translation [1]{[#1]}%
\providecommand \BibitemOpen [0]{}%
\providecommand \bibitemStop [0]{}%
\providecommand \bibitemNoStop [0]{.\EOS\space}%
\providecommand \EOS [0]{\spacefactor3000\relax}%
\providecommand \BibitemShut  [1]{\csname bibitem#1\endcsname}%
\let\auto@bib@innerbib\@empty
\bibitem [{\citenamefont {Suen}\ \emph {et~al.}(1992)\citenamefont {Suen},
  \citenamefont {Engel}, \citenamefont {Santos}, \citenamefont {Shayegan},\
  and\ \citenamefont {Tsui}}]{Suen.PRL.92}%
  \BibitemOpen
  \bibfield  {author} {\bibinfo {author} {\bibfnamefont {Y.~W.}\ \bibnamefont
  {Suen}}, \bibinfo {author} {\bibfnamefont {L.~W.}\ \bibnamefont {Engel}},
  \bibinfo {author} {\bibfnamefont {M.~B.}\ \bibnamefont {Santos}}, \bibinfo
  {author} {\bibfnamefont {M.}~\bibnamefont {Shayegan}}, \ and\ \bibinfo
  {author} {\bibfnamefont {D.~C.}\ \bibnamefont {Tsui}},\ }\href {\doibase
  10.1103/PhysRevLett.68.1379} {\bibfield  {journal} {\bibinfo  {journal}
  {Phys. Rev. Lett.}\ }\textbf {\bibinfo {volume} {68}},\ \bibinfo {pages}
  {1379} (\bibinfo {year} {1992})}\BibitemShut {NoStop}%
\bibitem [{\citenamefont {Manoharan}\ \emph {et~al.}(1997)\citenamefont
  {Manoharan}, \citenamefont {Suen}, \citenamefont {Lay}, \citenamefont
  {Santos},\ and\ \citenamefont {Shayegan}}]{Manoharan.PRL.97}%
  \BibitemOpen
  \bibfield  {author} {\bibinfo {author} {\bibfnamefont {H.~C.}\ \bibnamefont
  {Manoharan}}, \bibinfo {author} {\bibfnamefont {Y.~W.}\ \bibnamefont {Suen}},
  \bibinfo {author} {\bibfnamefont {T.~S.}\ \bibnamefont {Lay}}, \bibinfo
  {author} {\bibfnamefont {M.~B.}\ \bibnamefont {Santos}}, \ and\ \bibinfo
  {author} {\bibfnamefont {M.}~\bibnamefont {Shayegan}},\ }\href {\doibase
  10.1103/PhysRevLett.79.2722} {\bibfield  {journal} {\bibinfo  {journal}
  {Phys. Rev. Lett.}\ }\textbf {\bibinfo {volume} {79}},\ \bibinfo {pages}
  {2722} (\bibinfo {year} {1997})}\BibitemShut {NoStop}%
\bibitem [{\citenamefont {Luhman}\ \emph {et~al.}(2008)\citenamefont {Luhman},
  \citenamefont {Pan}, \citenamefont {Tsui}, \citenamefont {Pfeiffer},
  \citenamefont {Baldwin},\ and\ \citenamefont {West}}]{Luhman.PRL.08}%
  \BibitemOpen
  \bibfield  {author} {\bibinfo {author} {\bibfnamefont {D.~R.}\ \bibnamefont
  {Luhman}}, \bibinfo {author} {\bibfnamefont {W.}~\bibnamefont {Pan}},
  \bibinfo {author} {\bibfnamefont {D.~C.}\ \bibnamefont {Tsui}}, \bibinfo
  {author} {\bibfnamefont {L.~N.}\ \bibnamefont {Pfeiffer}}, \bibinfo {author}
  {\bibfnamefont {K.~W.}\ \bibnamefont {Baldwin}}, \ and\ \bibinfo {author}
  {\bibfnamefont {K.~W.}\ \bibnamefont {West}},\ }\href {\doibase
  10.1103/PhysRevLett.101.266804} {\bibfield  {journal} {\bibinfo  {journal}
  {Phys. Rev. Lett.}\ }\textbf {\bibinfo {volume} {101}},\ \bibinfo {pages}
  {266804} (\bibinfo {year} {2008})}\BibitemShut {NoStop}%
\bibitem [{\citenamefont {Shabani}\ \emph {et~al.}(2010)\citenamefont
  {Shabani}, \citenamefont {Liu},\ and\ \citenamefont
  {Shayegan}}]{Shabani.PRL.10}%
  \BibitemOpen
  \bibfield  {author} {\bibinfo {author} {\bibfnamefont {J.}~\bibnamefont
  {Shabani}}, \bibinfo {author} {\bibfnamefont {Y.}~\bibnamefont {Liu}}, \ and\
  \bibinfo {author} {\bibfnamefont {M.}~\bibnamefont {Shayegan}},\ }\href
  {\doibase 10.1103/PhysRevLett.105.246805} {\bibfield  {journal} {\bibinfo
  {journal} {Phys. Rev. Lett.}\ }\textbf {\bibinfo {volume} {105}},\ \bibinfo
  {pages} {246805} (\bibinfo {year} {2010})}\BibitemShut {NoStop}%
\bibitem [{\citenamefont {Manoharan}\ \emph {et~al.}(1996)\citenamefont
  {Manoharan}, \citenamefont {Suen}, \citenamefont {Santos},\ and\
  \citenamefont {Shayegan}}]{Manoharan.PRL.96}%
  \BibitemOpen
  \bibfield  {author} {\bibinfo {author} {\bibfnamefont {H.~C.}\ \bibnamefont
  {Manoharan}}, \bibinfo {author} {\bibfnamefont {Y.~W.}\ \bibnamefont {Suen}},
  \bibinfo {author} {\bibfnamefont {M.~B.}\ \bibnamefont {Santos}}, \ and\
  \bibinfo {author} {\bibfnamefont {M.}~\bibnamefont {Shayegan}},\ }\href
  {\doibase 10.1103/PhysRevLett.77.1813} {\bibfield  {journal} {\bibinfo
  {journal} {Phys. Rev. Lett.}\ }\textbf {\bibinfo {volume} {77}},\ \bibinfo
  {pages} {1813} (\bibinfo {year} {1996})}\BibitemShut {NoStop}%
\bibitem [{\citenamefont {Winkler}(2003)}]{Winkler.03}%
  \BibitemOpen
  \bibfield  {author} {\bibinfo {author} {\bibfnamefont {R.}~\bibnamefont
  {Winkler}},\ }\href@noop {} {\emph {\bibinfo {title} {Spin-Orbit Coupling
  Effects in Two-Dimensional Electron and Hole Systems}}}\ (\bibinfo
  {publisher} {Springer, Berlin},\ \bibinfo {year} {2003})\BibitemShut
  {NoStop}%
\bibitem [{\citenamefont {Lay}\ \emph {et~al.}(1997)\citenamefont {Lay},
  \citenamefont {Jungwirth}, \citenamefont {Smr\ifmmode~\check{c}\else
  \v{c}\fi{}ka},\ and\ \citenamefont {Shayegan}}]{Lay.PRB.97}%
  \BibitemOpen
  \bibfield  {author} {\bibinfo {author} {\bibfnamefont {T.~S.}\ \bibnamefont
  {Lay}}, \bibinfo {author} {\bibfnamefont {T.}~\bibnamefont {Jungwirth}},
  \bibinfo {author} {\bibfnamefont {L.}~\bibnamefont
  {Smr\ifmmode~\check{c}\else \v{c}\fi{}ka}}, \ and\ \bibinfo {author}
  {\bibfnamefont {M.}~\bibnamefont {Shayegan}},\ }\href {\doibase
  10.1103/PhysRevB.56.R7092} {\bibfield  {journal} {\bibinfo  {journal} {Phys.
  Rev. B}\ }\textbf {\bibinfo {volume} {56}},\ \bibinfo {pages} {R7092}
  (\bibinfo {year} {1997})}\BibitemShut {NoStop}%
\bibitem [{\citenamefont {Hu}\ and\ \citenamefont
  {MacDonald}(1992)}]{Hu.PRB.92}%
  \BibitemOpen
  \bibfield  {author} {\bibinfo {author} {\bibfnamefont {J.}~\bibnamefont
  {Hu}}\ and\ \bibinfo {author} {\bibfnamefont {A.~H.}\ \bibnamefont
  {MacDonald}},\ }\href {\doibase 10.1103/PhysRevB.46.12554} {\bibfield
  {journal} {\bibinfo  {journal} {Phys. Rev. B}\ }\textbf {\bibinfo {volume}
  {46}},\ \bibinfo {pages} {12554} (\bibinfo {year} {1992})}\BibitemShut
  {NoStop}%
\bibitem [{\citenamefont {Murphy}\ \emph {et~al.}(1994)\citenamefont {Murphy},
  \citenamefont {Eisenstein}, \citenamefont {Boebinger}, \citenamefont
  {Pfeiffer},\ and\ \citenamefont {West}}]{Murphy.PRL.94}%
  \BibitemOpen
  \bibfield  {author} {\bibinfo {author} {\bibfnamefont {S.~Q.}\ \bibnamefont
  {Murphy}}, \bibinfo {author} {\bibfnamefont {J.~P.}\ \bibnamefont
  {Eisenstein}}, \bibinfo {author} {\bibfnamefont {G.~S.}\ \bibnamefont
  {Boebinger}}, \bibinfo {author} {\bibfnamefont {L.~N.}\ \bibnamefont
  {Pfeiffer}}, \ and\ \bibinfo {author} {\bibfnamefont {K.~W.}\ \bibnamefont
  {West}},\ }\href {\doibase 10.1103/PhysRevLett.72.728} {\bibfield  {journal}
  {\bibinfo  {journal} {Phys. Rev. Lett.}\ }\textbf {\bibinfo {volume} {72}},\
  \bibinfo {pages} {728} (\bibinfo {year} {1994})}\BibitemShut {NoStop}%
\bibitem [{\citenamefont {Fischer}\ \emph {et~al.}(2007)\citenamefont
  {Fischer}, \citenamefont {Winkler}, \citenamefont {Schuh}, \citenamefont
  {Bichler},\ and\ \citenamefont {Grayson}}]{Fischer.PRB.07}%
  \BibitemOpen
  \bibfield  {author} {\bibinfo {author} {\bibfnamefont {F.}~\bibnamefont
  {Fischer}}, \bibinfo {author} {\bibfnamefont {R.}~\bibnamefont {Winkler}},
  \bibinfo {author} {\bibfnamefont {D.}~\bibnamefont {Schuh}}, \bibinfo
  {author} {\bibfnamefont {M.}~\bibnamefont {Bichler}}, \ and\ \bibinfo
  {author} {\bibfnamefont {M.}~\bibnamefont {Grayson}},\ }\href {\doibase
  10.1103/PhysRevB.75.073303} {\bibfield  {journal} {\bibinfo  {journal} {Phys.
  Rev. B}\ }\textbf {\bibinfo {volume} {75}},\ \bibinfo {pages} {073303}
  (\bibinfo {year} {2007})}\BibitemShut {NoStop}%
\bibitem [{Note1()}]{Note1}%
  \BibitemOpen
  \bibinfo {note} {Both the multi-band structure of 2D holes and a tilted field
  geometry imply complicated couplings between different Landau harmonic
  oscillators. A joint treatment of both effects would be exorbitantly complex
  to implement and also demanding when solved numerically.}\BibitemShut {Stop}%
\bibitem [{\citenamefont {Davies}\ \emph {et~al.}(1991)\citenamefont {Davies},
  \citenamefont {Newbury}, \citenamefont {Pepper}, \citenamefont {Frost},
  \citenamefont {Ritchie},\ and\ \citenamefont {Jones}}]{Davies.PRB.91}%
  \BibitemOpen
  \bibfield  {author} {\bibinfo {author} {\bibfnamefont {A.~G.}\ \bibnamefont
  {Davies}}, \bibinfo {author} {\bibfnamefont {R.}~\bibnamefont {Newbury}},
  \bibinfo {author} {\bibfnamefont {M.}~\bibnamefont {Pepper}}, \bibinfo
  {author} {\bibfnamefont {J.~E.~F.}\ \bibnamefont {Frost}}, \bibinfo {author}
  {\bibfnamefont {D.~A.}\ \bibnamefont {Ritchie}}, \ and\ \bibinfo {author}
  {\bibfnamefont {G.~A.~C.}\ \bibnamefont {Jones}},\ }\href {\doibase
  10.1103/PhysRevB.44.13128} {\bibfield  {journal} {\bibinfo  {journal} {Phys.
  Rev. B}\ }\textbf {\bibinfo {volume} {44}},\ \bibinfo {pages} {13128}
  (\bibinfo {year} {1991})}\BibitemShut {NoStop}%
\bibitem [{\citenamefont {Muraki}\ and\ \citenamefont
  {Hirayama}(1999)}]{Muraki.PRB.99}%
  \BibitemOpen
  \bibfield  {author} {\bibinfo {author} {\bibfnamefont {K.}~\bibnamefont
  {Muraki}}\ and\ \bibinfo {author} {\bibfnamefont {Y.}~\bibnamefont
  {Hirayama}},\ }\href {\doibase 10.1103/PhysRevB.59.R2502} {\bibfield
  {journal} {\bibinfo  {journal} {Phys. Rev. B}\ }\textbf {\bibinfo {volume}
  {59}},\ \bibinfo {pages} {R2502} (\bibinfo {year} {1999})}\BibitemShut
  {NoStop}%
\bibitem [{\citenamefont {Pan}\ \emph {et~al.}(2005)\citenamefont {Pan},
  \citenamefont {Cs\'athy}, \citenamefont {Tsui}, \citenamefont {Pfeiffer},\
  and\ \citenamefont {West}}]{Pan.PRB.05}%
  \BibitemOpen
  \bibfield  {author} {\bibinfo {author} {\bibfnamefont {W.}~\bibnamefont
  {Pan}}, \bibinfo {author} {\bibfnamefont {G.~A.}\ \bibnamefont {Cs\'athy}},
  \bibinfo {author} {\bibfnamefont {D.~C.}\ \bibnamefont {Tsui}}, \bibinfo
  {author} {\bibfnamefont {L.~N.}\ \bibnamefont {Pfeiffer}}, \ and\ \bibinfo
  {author} {\bibfnamefont {K.~W.}\ \bibnamefont {West}},\ }\href {\doibase
  10.1103/PhysRevB.71.035302} {\bibfield  {journal} {\bibinfo  {journal} {Phys.
  Rev. B}\ }\textbf {\bibinfo {volume} {71}},\ \bibinfo {pages} {035302}
  (\bibinfo {year} {2005})}\BibitemShut {NoStop}%
\end{thebibliography}
\end{document}